\documentclass[fleqn,usenatbib]{mnras}

\usepackage{newtxtext,newtxmath}

\usepackage[T1]{fontenc}
\usepackage{ae,aecompl}

\usepackage{graphicx}	
\usepackage{amsmath}	
\usepackage{amssymb}	

\newcommand{\kms}{\mbox{km\,s$^{-1}$}}

\newcommand{\fwhm}{\hbox{$\Delta$v$_{\tiny{\textrm{FWHM}}}$}}

\newcommand{\NX}{\hbox{$N_{H_2}^{12}$}}
\newcommand{\Nthir}{\hbox{$N_{H_2}^{13}$}}

\newcommand{\RXJ}{\hbox{RX\,J1713.7$-$3946}}
\newcommand{\Jsto}{\hbox{HESS\,J1731$-$347}}

\newcommand{\Jftf}{\hbox{HESS\,J1534$-$571}}
\newcommand{\VJ}{\hbox{RX\,J0852.0$-$4622}}

\newcommand{\HI}{\hbox{H\,{\sc i}}}

\newcommand{\cmsqr}{\hbox{cm$^{-2}$}}

\newcommand{\Msun}{\hbox{M$_{\odot}$}}
\newcommand{\miriad}{\hbox{\sc Miriad}}

\title[An ISM association for HESS\,J1534$-$571]{Searching for an interstellar medium association for HESS\,J1534$-$571}

\author[N. I. Maxted et al.]{Nigel I. Maxted$^{1,2}$\thanks{Email: n.maxted@unsw.edu.au}, 
C. Braiding,$^{1}$
G. F. Wong,$^{1,2}$ 
G. P. Rowell,$^{3}$ 
M. G. Burton,$^{1,4}$ \newauthor
M. D. Filipovi\'c,$^{2}$ 
F. Voisin,$^{3}$
D. Uro\v{s}evi\'c,$^{5,6}$ 
B. Vukoti\'c,$^{7}$
M. Z. Pavlovi\'c,$^{5}$ 
H. Sano$^{8}$ \newauthor
and Y. Fukui$^{8}$\\ 
$^{1}$School of Physics, The University of New South Wales, Sydney, 2052, Australia\\
$^2$Western Sydney University, Locked Bag 1797, Penrith, NSW 2751, Australia\\
$^3$School of Physical Sciences, The University of Adelaide, Adelaide, 5005,  Australia\\
$^4$Armagh Observatory and Planetarium, College Hill, Armagh, BT61 9DG, Northern Ireland, United Kingdom\\
$^5$Department of Astronomy Faculty of Mathematics, University of Belgrade Studentski trg 16, 11000 Belgrade, Serbia\\
$^6$Isaac Newton Institute of Chile Yugoslavia Branch, Serbia\\
$^7$Astronomical Observatory, Volgina 7, P.O.Box 74 11060 Belgrade, Serbia\\
$^8$Department of Physics, Nagoya University, Furo-cho, Chikusa-ku, Nagoya 464-8601, Japan
}

\date{Accepted 2018 July 03. Received 2018 July 01; in original form 2018 April 27}

\pubyear{2018}

\begin{document}
\label{firstpage}
\pagerange{\pageref{firstpage}--\pageref{lastpage}}
\maketitle

\begin{abstract}
The Galactic supernova remnant $\Jftf$ (also known as G323.7$-$1.0) has a shell-like morphology in TeV gamma-ray emission and is a key object in the study of cosmic ray origin. 
Little is known about its distance and local environment. We examine Mopra $^{12}$CO/$^{13}$CO(1-0) data, Australian Telescope Compact Array $\HI$, and Parkes $\HI$ data towards $\Jftf$. We trace molecular clouds in at least five velocity ranges, including clumpy interstellar medium structures near a dip in $\HI$ emission at a kinematic velocity consistent with the Scutum-Crux arm at $\sim$3.5\,kpc. This feature may be a cavity blown-out by the progenitor star, a scenario that suggests $\Jftf$ resulted from a core-collapse event. By employing parametrisations fitted to a sample of supernova remnants of known distance, we find that the radio continuum brightness of $\Jftf$ is consistent with the $\sim$3.5\,kpc kinematic distance of the Scutum-Crux arm $\HI$ dip. Modelling of the supernova evolution suggests an $\sim$8-24\,kyr age for $\Jftf$ at this distance. 
\end{abstract}





\section{Introduction}
\begin{figure}
\begin{center}
\includegraphics[width=0.48\textwidth, angle=0,trim={0 0 0 0},clip]{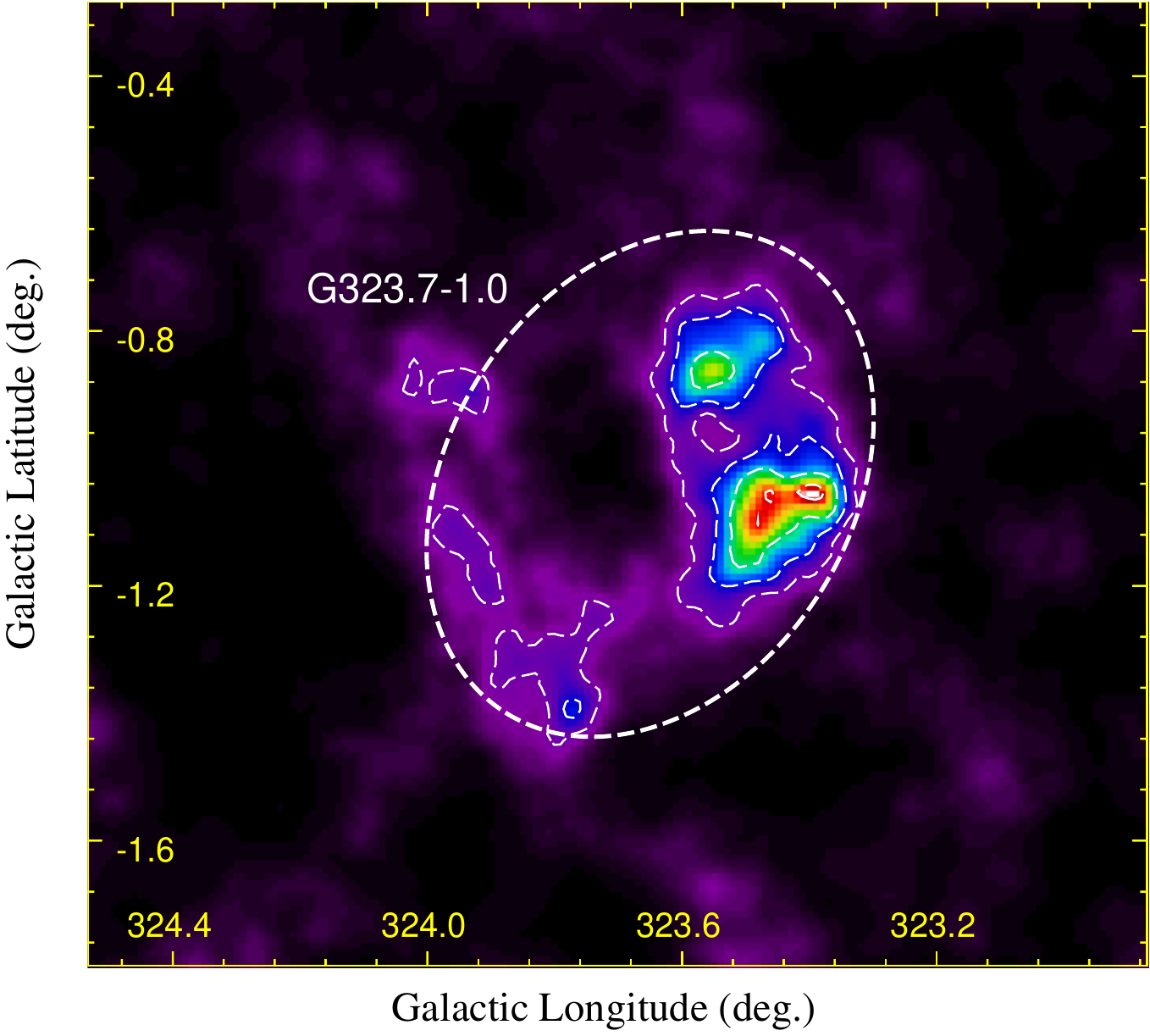}
\caption{High Energy Stereoscopic System (HESS) $\sim$0.6-100\,TeV gamma-ray significance image of $\Jftf$ \citep{Abramowski:2017newshells}. Corresponding 3, 4, 5 and 6$\sigma$ TeV gamma-ray emission contours are overlaid. A white-dashed ellipse indicates the position of SNR G323.7$-$1.0 \citep{Green:2014}, which was first identified in 843\,MHz radio continuum emission from the Molonglo Observatory Synthesis telescope (MOST) survey. The ellipse has a centre of [$\alpha$,$\delta$, J2000]=[05:34:30.1$-$57:12:03] and axes diameter of 51$^{\prime}\times$38$^{\prime}$. \label{fig:HESS}}
\end{center}
\end{figure}

The Galactic supernova remnant candidate G323.7$-$1.0 was initially discovered in 843\,MHz radio continuum emission \citep[see][]{Green:2014}. Its nature as a shell-type supernova remnant (SNR) was confirmed with the significant detection of a corresponding TeV gamma-ray shell, $\Jftf$, as shown in Figure\,\ref{fig:HESS} \citep{Abramowski:2017newshells}. Supernova remnants with associated gamma-ray emission are key targets in the search for Galactic sources of cosmic ray hadrons (hereafter CRs). Supernova shocks with velocities $>$1000\,km\,s$^{-1}$ may be able to host the diffusive shock acceleration process for particle acceleration \citep[e.g.][]{Bell:1978}, which would help to explain the observed flux of Galactic CRs up to energies of several PeV. 

Some mature ($\sim$10$^{4}$\,yr) shell-type SNRs have been proven to be CR accelerators \citep[e.g. IC\,443 and W44,][]{Ackermann:2013} through observations of high energy spectra which have so-called `pion-bumps' - signatures of neutral-pion creation and decay resulting from CR interactions with gas (i.e. a proton-proton mechanism). Some SNRs have a spatial correspondence between gas and gamma-rays that strongly suggests this mechanism 
\citep[e.g. W28 and W49,][]{Aharonian:w28,Abdalla:2016w49}. 
In these cases, it is believed that the SNR shocks have created the conditions needed to accelerate protons to TeV energies. These protons (TeV-energy CR hadrons) have then diffused into nearby molecular clouds where proton-proton interactions lead to observable gamma-ray emission that reflects the morphology of the associated gas. Corresponding populations of high-energy electrons, also accelerated by the SNR shocks, have cooled to sub-TeV energies more quickly than protons in this scenario due to the synchrotron emission mechanism. 
The growing list of SNRs with strong ties to TeV CRs tells a convincing story about the role of these objects in CR acceleration at TeV energies. 

A sub-class of gamma-ray SNRs with shell-type morphology at TeV energies is comprised by eight objects - $\RXJ$, $\Jsto$, $\VJ$, RCW\,86, SN\,1006 and three new SNRs and SNR candidates, HESS\,J1614$-$518, HESS\,J1912$+$101 and $\Jftf$. The former 5 of this group (generally aged $\sim$10$^3$\,yr), have a GeV-TeV spectral shape that appears to favour a gamma-ray production process that involves high-energy electrons \citep[e.g.][]{Acero:2015} - inverse-Compton scattering of low-energy photons (leptonic emission), within or near the SNR shock. However,  
it has been argued that a clumpy interstellar medium might be able to alter the expected hadronic gamma-ray spectral shape of young TeV SNRs if energy-dependent diffusion into dense gas clumps within or near the SNR shock front plays a significant role \citep{Gabici:2009,Zirakashvili:2010,Inoue:2012,Fukui:2012,Maxted:2012,Gabici:2014}. Evidence of a gamma-ray production mechanism inherently linked to gas density is present in the strong correspondence between gas proton density and gamma-ray flux in $\RXJ$, $\Jsto$ and $\VJ$ \citep[][respectively]{Fukui:2012,Fukuda:2014,Fukui:2017}. The authors point to the hadronic gamma-ray production mechanism as the likely explanation. In this respect, knowledge of the ISM local to SNRs is important for understanding the nature of the gamma-ray production in individual cases. 

Gas associations for young ($\sim$10$^3$\,yr) shell-type SNRs have been difficult to prove definitively. Older ($\sim$10$^4$\,yr) SNRs may exhibit signatures of shocked gas that have corresponding measurable kinematic distances, e.g. observations of spectral line broadening or shock-tracing SiO or 1720\,MHz OH emission \citep[e.g.][respectively]{Nicholas:2012,Frail:1996}. In contrast, the high-speed shocks ($>$1000\,kms$^{-1}$) of young Galactic shell-type SNRs have not yet begun to exhibit clear markers of shocked gas that contain line of sight velocity information. In such cases, gas associations are sometimes inferred using morphological correspondence between X-rays and gamma-rays, and $\HI$ and CO. This is the case for SNRs $\RXJ$, $\Jsto$ and $\VJ$ \citep[][]{Fukui:2012,Fukuda:2014,Fukui:2017}. 

$\Jftf$ may share some similarities with $\RXJ$, $\Jsto$ and $\VJ$, both morphologically and spectrally. \citet{Araya:2017} recently argued for a leptonic scenario in $\Jftf$, following a GeV gamma-ray spectral analysis. In this scenario, a distance of $\sim$4-5\,kpc was inferred by the level of far infrared emission required to produce the observed gamma-ray emission via an assumed Inverse-Compton scattering scenario. 
Although initial attempts to find a SNR-association in Suzaku X-ray data revealed no significant counterpart \citep{Puehlhofer:2016,Abramowski:2017newshells}, regions of soft 0.5-3\,keV X-rays towards $\Jftf$ were later discovered, while 6.4\,keV Fe-line emission suggested the existence of an ionising population of $\sim$10\,MeV-energy CR protons \citep{Saji:2018}. Soft X-ray absorption modelling suggested a SNR at distance of $\sim$6$\pm$2\,kpc \citep{Saji:2018}, in agreement with the \citeauthor{Araya:2017} distance ($\sim$4-5\,kpc, assuming leptonic gamma-ray emission). A specific gas association at such distances has not been identified, however one is suggested in our study.

The local environment of individual SNRs (among other parameters) can also have an effect on the robustness of the radio surface brightness to diameter relation \citep[e.g.][]{Kostic:2016}, i.e $\Sigma$-$D$ relation, particularly in the case of shock-cloud interactions \citep[see][for detailed modelling]{Pavlovic:2017}. In the case of $\Jftf$, which is suggested to be under-luminous at radio wavelengths, \citet{Abramowski:2017newshells} scaled the $\Sigma$-$D$ relation to match the case of the shell-type SNR $\RXJ$, which is also known to be radio-dim. The resultant inferred distance was $\sim$5\,kpc, consistent with aforementioned values \citep[6$\pm$2 and $\sim$4-5\,kpc][]{Araya:2017,Saji:2018}. This may imply some similarities between $\RXJ$ and $\Jftf$, and the ISM local to both. $\RXJ$ exists in a particularly clumpy medium, but the environment local to $\Jftf$ has not been examined at a sufficient angular resolution until now. 

The identification of a gas association for $\Jftf$ will help to further constrain the distance and pinpoint potential CR target material related to gamma-ray emission. For this purpose, Mopra CO(1-0) data and SGPS $\HI$ data are employed to investigate the ISM towards this new shell-type gamma-ray SNR. We particularly target molecular clumps.

We outline the characteristics of our data and analysis in Section\,\ref{sec:obs}. We present new Mopra maps of CO(1-0) emission towards $\Jftf$ in Section\,\ref{sec:results}. In Section\,\ref{sec:Disc} we identify molecular clump features consistent with an association with $\Jftf$ before discussing some implications for the implied distance. The association is reinforced by a statistical study of the radio flux of $\Jftf$ compared to a sample of SNRs.

\section{Observations}\label{sec:obs}
In this study, we utilise Mopra CO Galactic Plane Survey data to identify molecular clumps in the line of sight of the Galactic SNR $\Jftf$. Mopra spectral CO(1-0) data has a superior angular resolution to previously-published gas data towards the $\Jftf$ region. Public Southern Galactic Plane Survey (SGPS) $\HI$ data is re-examined \citep[following studies by][]{Abramowski:2017newshells} in support of this new Mopra data-set. Where relevant to interpreting multi-wavelength data, molecular column densities and masses are calculated.

\subsection{The Mopra CO Galactic Plane Survey at 3 mm}
Spectral data of the J$=$1-0 transition of CO isotopologues was taken as part of the Mopra Galactic Plane CO Survey\footnote{http://phys.unsw.edu.au/mopraco/} \citep[see][]{Burton:2013}\footnote{see www.mopra.org/data/}. The $^{12}$CO\footnote{hereafter CO}, $^{13}$CO, C$^{18}$O and C$^{17}$O(1-0) were targeted in the Mopra survey, but only the former two of these isotopologues are exploited in this paper, as they are the brightest emitters.

Details of the latest Mopra CO data processing procedures are presented in \citet{Braiding:2015} and \citet{Braiding:2018}. Data towards the $\Jftf$ field have a velocity coverage of $-$596\,kms$^{-1}$ $<$v$_{LSR}<+$501\,kms$^{-1}$ and $-$524\,kms$^{-1}$ $<$v$_{LSR}<+$247\,kms$^{-1}$ for CO(1-0) and $^{13}$CO(1-0), respectively. The angular resolution and velocity resolution are 35$^{\prime\prime}$ and 0.1\,km\,s$^{-1}$, respectively, across eight 4096-channel dual-polarisation bands. The extended beam efficiency used to recalibrate the antenna temperature of Mopra at 115\,GHz is 0.55 \citep{Ladd:2005}.

\subsection{21 cm SGPS HI data}
Publicly-available 21\,cm $\HI$ data from the SGPS \citep[see][ for details]{McClure:2005} was examined in support of our CO analysis (see Appendix\,\ref{app:HI}). $\HI$ fits cubes were created by combining Parkes single-dish and Australia Telescope Compact Array (ATCA) interferometer data to cover both large and small-scale structures of atomic gas with a resultant angular and velocity resolution of $\sim$2$^{\prime}$ and $\sim$0.8\,$\kms$, respectively. $\HI$ data cubes have been calibrated to remove radio-continuum sources.

\subsection{Spectral line examination}
\begin{figure*}
\begin{center}
\includegraphics[width=0.81\textwidth]{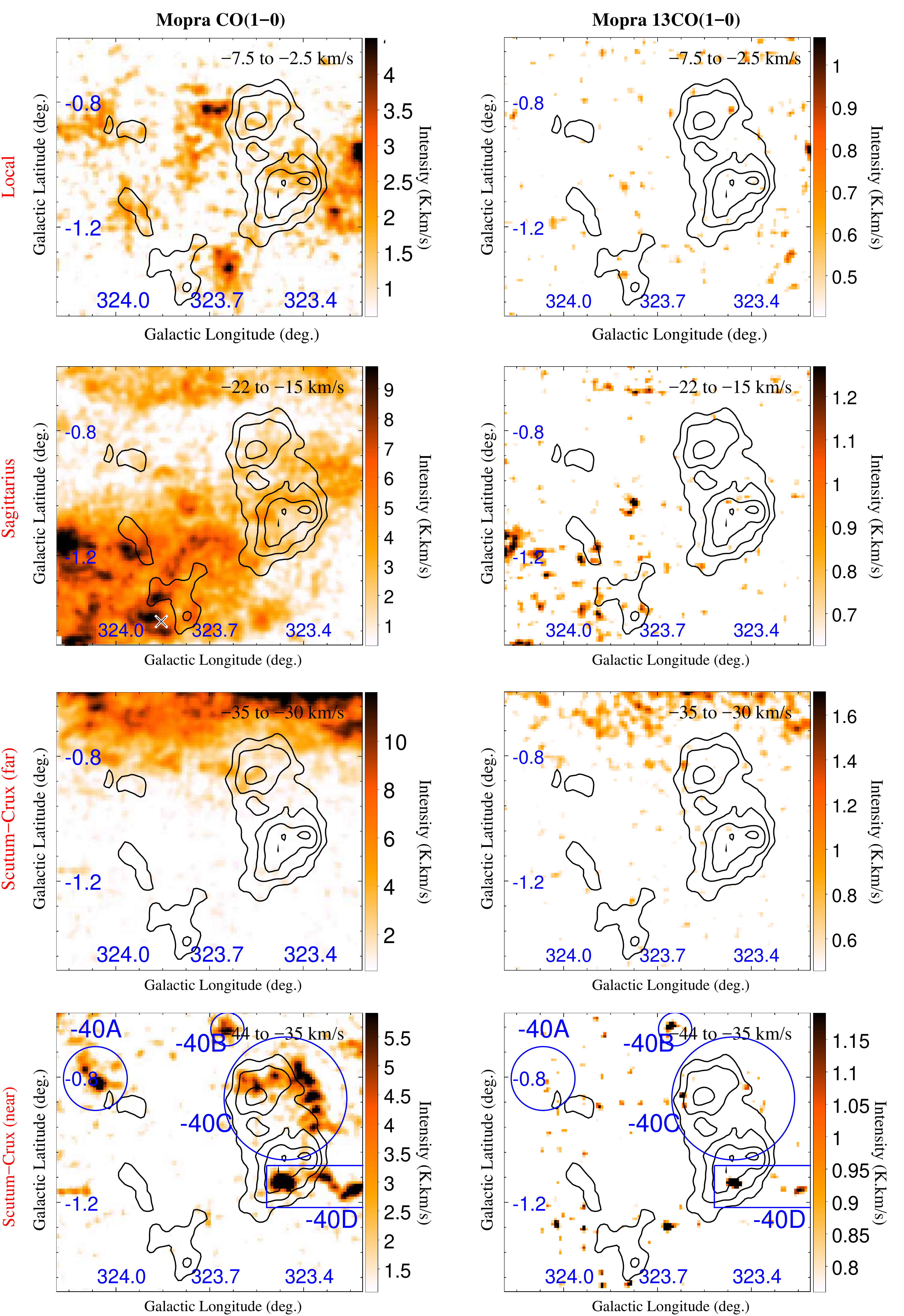}
\caption{Velocity integrated CO(1-0) (left) and $^{13}$CO(1-0) (right) images. Integration ranges which are indicated inset on each image, are $-$7.5 to $-$2.5, $-$22 to $-$15, $-$35 to $-$30 and $-$44 to $-$35\,$\kms$, from top to bottom, respectively.  HESS 3, 4, 5 and 6$\sigma$ $>$1\,TeV gamma-ray significance contours (thick black) are overlaid \citep{Abramowski:2017newshells}. Suggested Galactic arm associations are indicated in red on along the left side of the figure for each row. In the bottom images, circular regions corresponding to parameter calculations in Table\,\ref{tab:clumps} are indicated.}\label{fig:Mom0s_1}
\end{center}
\end{figure*}

\begin{figure*}
\begin{center}
\includegraphics[width=0.81\textwidth]{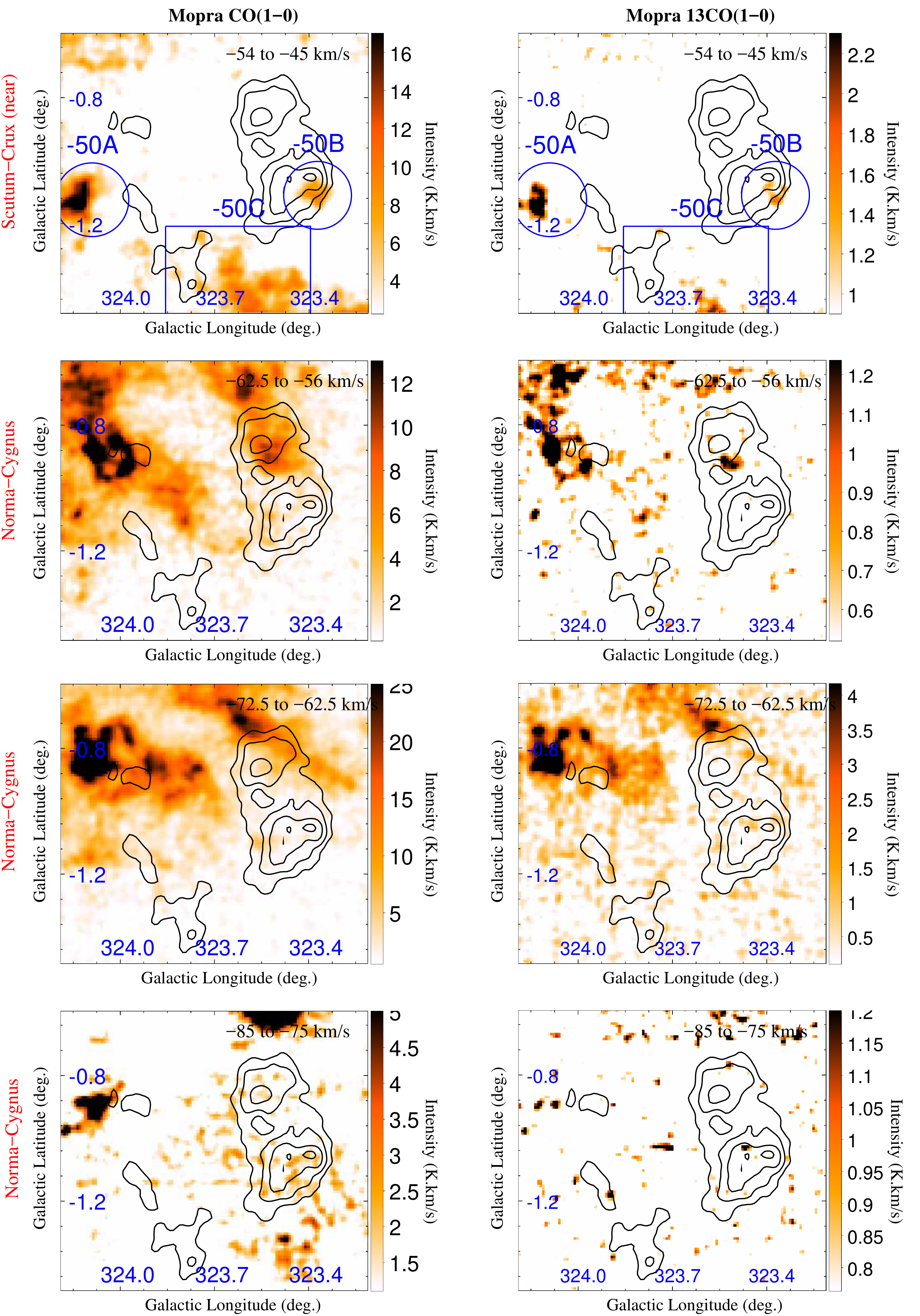}
\caption{Velocity integrated CO(1-0) (left) and $^{13}$CO(1-0) (right) images. Integration ranges which are indicated inset on each image, are $-$55.4 to $-$45, $-$62.5 to $-$56, $-$72.5 to $-$62.5 and $-$85 to $-$75\,$\kms$, from top to bottom, respectively.  HESS 3, 4, 5 and 6$\sigma$ $>$1\,TeV gamma-ray significance contours (thick black) are overlaid \citep{Abramowski:2017newshells}. Suggested Galactic arm associations are indicated in red on along the left side of the figure for each row. In the top images, circular regions corresponding to parameter calculations in Table\,\ref{tab:clumps} are indicated.}\label{fig:Mom0s_2}
\end{center}
\end{figure*}

We examine Mopra CO and $^{13}$CO(1-0) emission to identify molecular cloud components that might be candidates for association with $\Jftf$. Mopra CO(1-0) and $^{13}$CO(1-0) integrated intensity images are shown in Figures\,\ref{fig:Mom0s_1} and \ref{fig:Mom0s_2}. Velocity-integration ranges were chosen to highlight individual spectral features identified in the CO(1-0) cube. $\miriad$ software was used for this task \citep{Sault:1995}. 

\begin{figure*}
\begin{center}
\includegraphics[width=0.85\textwidth]{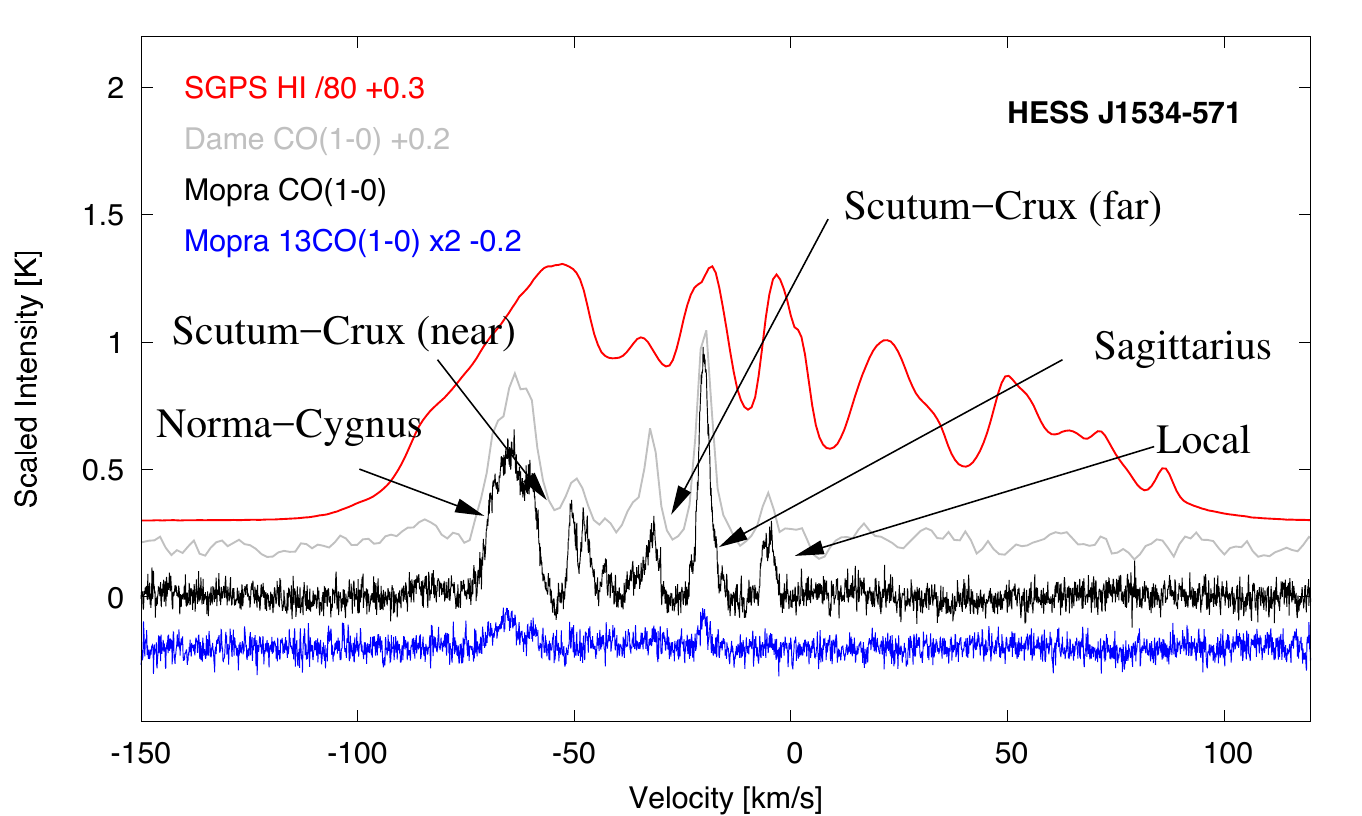}
\caption{Mopra CO, Mopra $^{13}$CO and SGPS \HI\ emission towards the elliptical MOST 843\,MHz emission region encompassing $\Jftf$, as displayed in Figure\,\ref{fig:HESS}. \label{fig:Spectra}}
\end{center}
\end{figure*}

\begin{figure*}
\begin{center}
a) Mopra CO(1-0) \hspace{5.8cm} b) Mopra $^{13}$CO(1-0)\\
\includegraphics[width=0.45\textwidth]{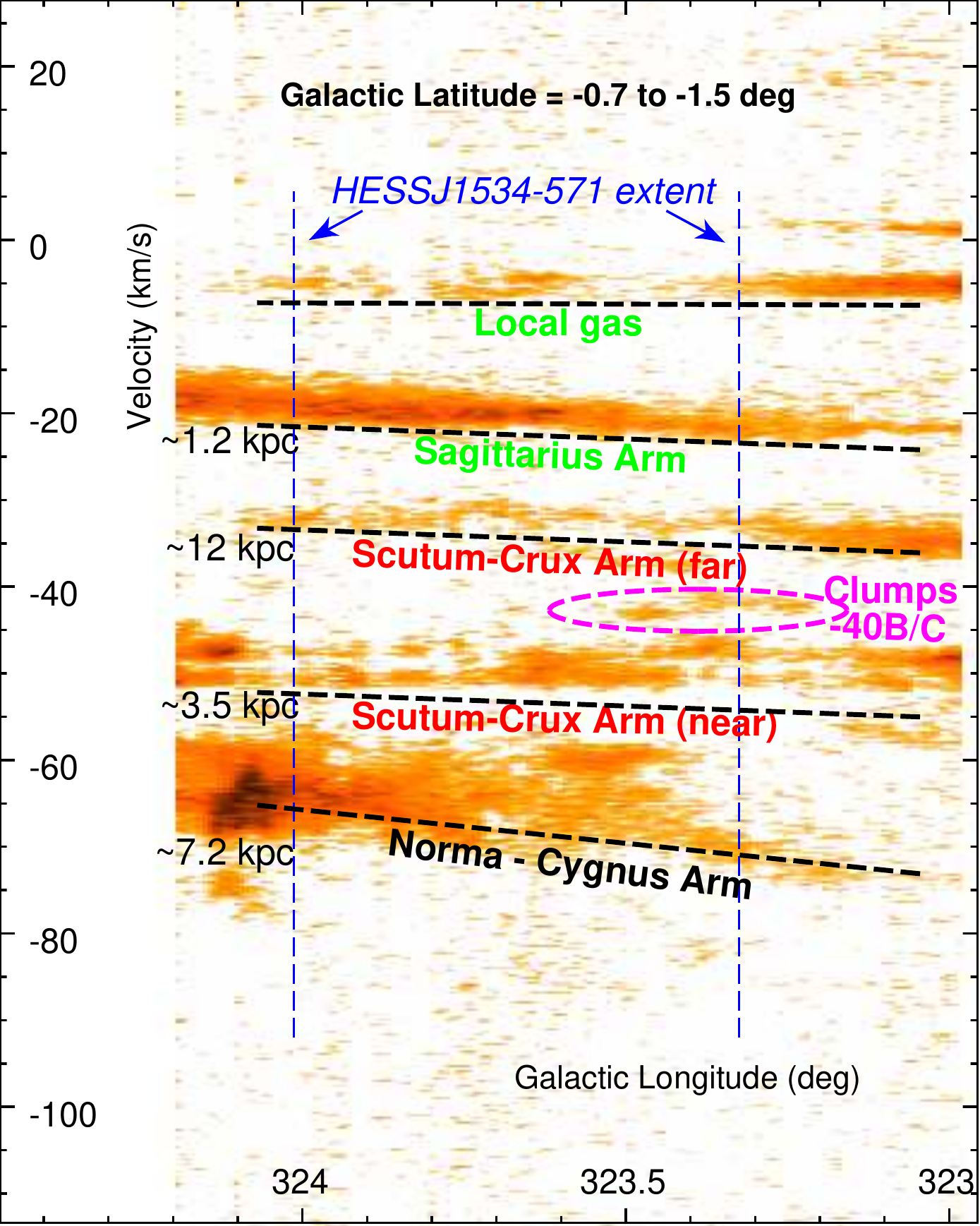}
\includegraphics[width=0.45\textwidth]{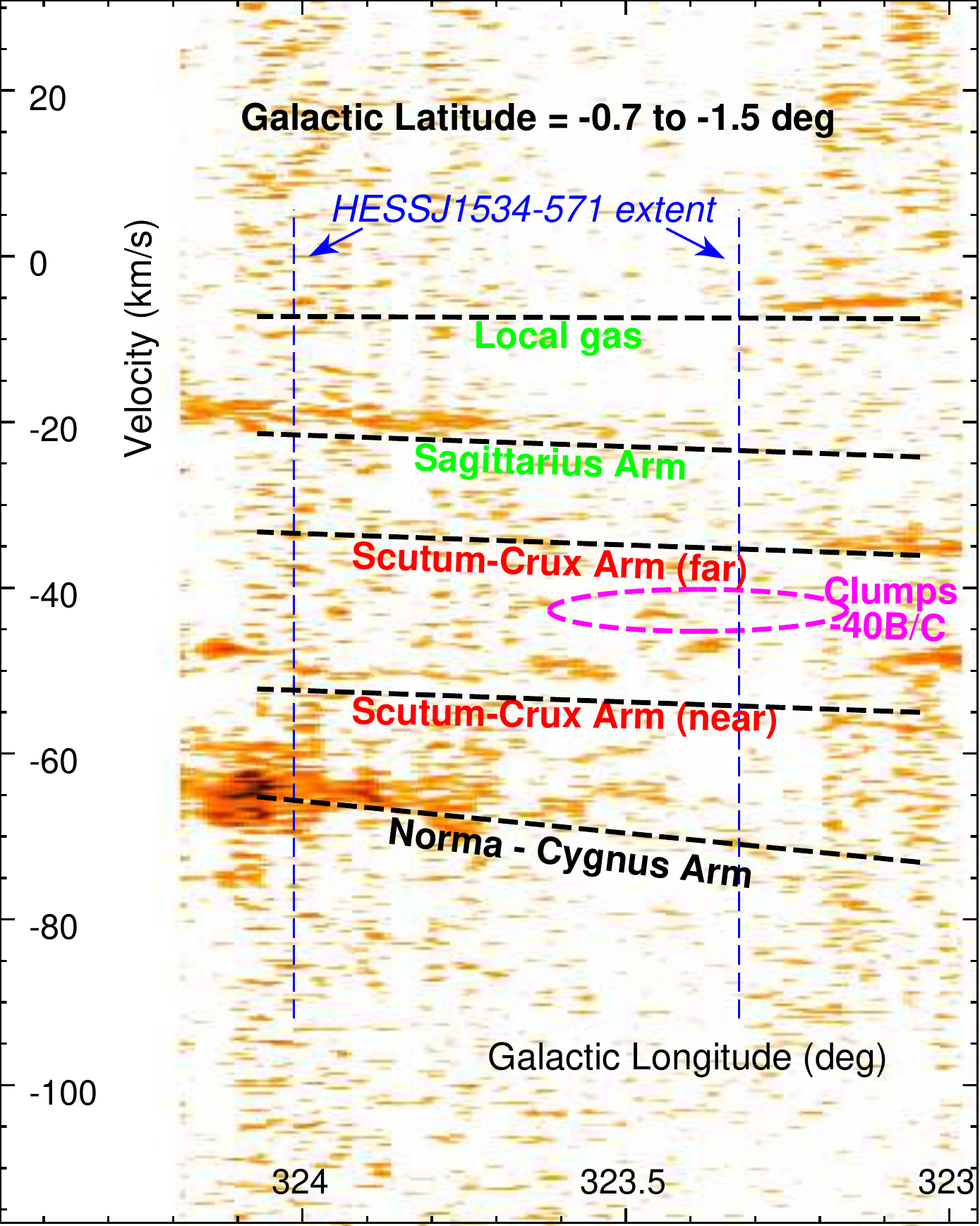}\\
c) SGPS \HI\ \hspace{6.8cm} d) The Milky Way Schematic\\
\includegraphics[width=0.45\textwidth]{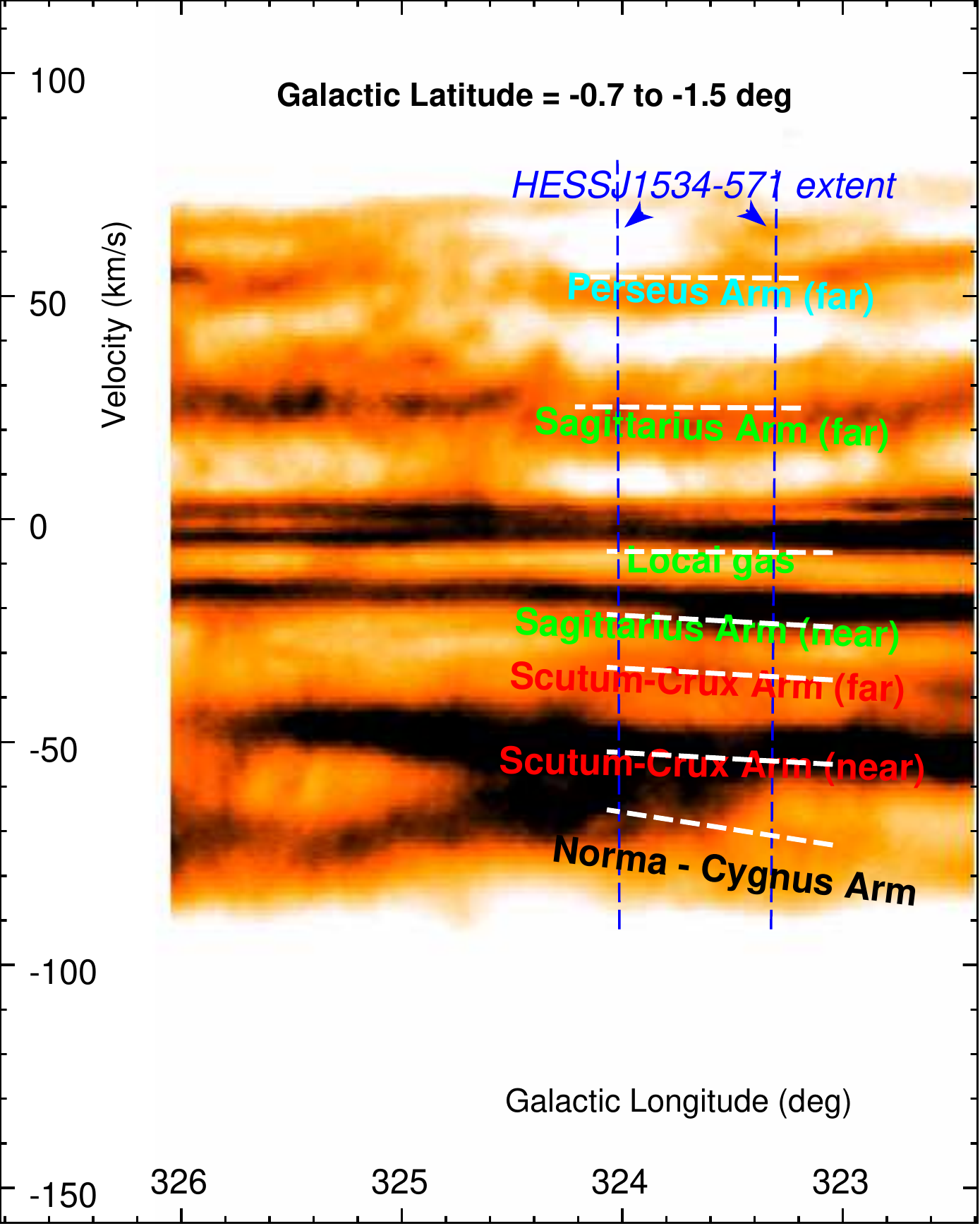}
\includegraphics[width=0.54\textwidth,trim={0cm 0cm 1.2cm 0.6cm},clip]{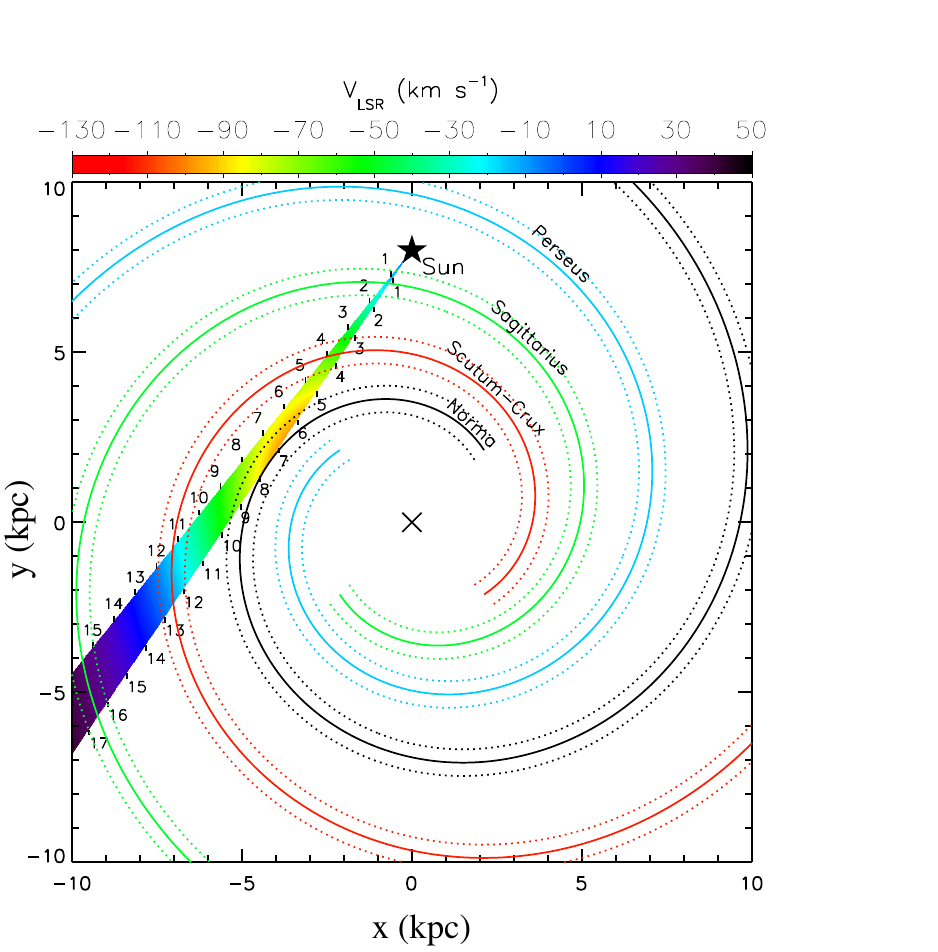}
\caption{Plots of Mopra CO(1-0) (a, top-left), Mopra $^{13}$CO(1-0) (b, top-right) and SGPS \HI\ (c, bottom) emission as a function of line of sight velocity and Galactic Longitude. CO, $^{13}$CO and \HI\ emission has been spatially integrated between $-$1.5 and $-$0.7$^{\circ}$. In images a, b and c, blue dotted lines indicate the longitudinal extent of $\Jftf$, and plausible Galactic arm associations are stated (see d). These lines correspond to the same coordinates in each image. Clumps $-40$B/C are indicated by ovals in figures a and b. d) A diagram of the Galactic rotation model of the Milky Way Galaxy \citep[see][for details]{Vallee:2016}. Based on (d), distances to corresponding arm features are shown in (a). \label{fig:PVplots}}
\end{center}
\end{figure*}

Figure\,\ref{fig:Spectra} and \ref{fig:PVplots}a, b and c display the velocity structure of the CO(1-0), $^{13}$CO(1-0) and \HI\ emission, respectively. Galactic arm structures are discernible (discussed in Section\,\ref{sec:results}) and consistent with that of the Columbia CO survey \citep{Dame:2001}, which is also shown in Figure\,\ref{fig:Spectra}. Figure\,\ref{fig:PVplots}d is used as a guide for the conversion between kinematic velocity and kinematic distance.

\subsection{Parameter calculations}
\subsubsection{$^{12}$CO Analysis}
Column densities and masses were calculated for CO cores identified in this analysis. An X-factor was used to convert CO(1-0) emission into H$_2$ column density, $N=X\int I_v dv$~~[cm$^{-2}$(K\,$\kms$)$^{-1}$], where $I_v dv$ is the velocity-integrated emission, $X$ is the X-factor and $N$ is the molecular column density \citep[e.g.][]{Lang:1980}. 
In our analysis, we use a range of CO X-factors, 1-3$\times 10^{20}\mathrm{cm}^{-2}.\mathrm{(Kkms}^{-1})^{-1}$ to account for the observed scatter of this value \citep[e.g.][]{Bolatto:2013,Okamoto:2017}.

\subsubsection{$^{13}$CO Analysis}
Where the less-abundant $^{13}$CO(1-0) transition is present, we calculate the CO optical depth to account for the attenuation in CO(1-0) via the prescription in \citet{Wilson:2013}. With the assumption that $^{13}$CO and $^{12}$CO emission are emitted from the same uniform excitation temperature region, we create excitation temperature maps via
\begin{equation}
\label{equ:Tex}
T=\frac{5.5}{\ln{\left(1+\dfrac{5.5}{T^{12}_{\rm B}+0.82}\right)}}~~~~~\textrm{[K]}
\end{equation} which is then used to calculate the $^{13}$CO(1-0) optical depth,
\begin{equation}
\label{equ:tau13}
\tau ^{13} _0 = -\ln{ \left[ 1 - \frac{T^{13}_{\rm B}}{5.3} \left[ \left( \exp{\left[ \frac{5.3}{T} \right] }-1 \right)^{-1} -0.16 \right] ^{-1} \right]  }
\end{equation} where T$_{\rm B} ^{12}$ and T$_B ^{\rm 13}$ are the $^{12}$CO and $^{13}$CO brightness temperatures, respectively. Equation\,\ref{equ:tau13} is valid under the assumption that the $^{12}$CO(1-0) optical depth is large compared to unity and the $^{13}$CO(1-0) optical depth is small compared to unity. The $^{13}$CO column density was then calculated by
\begin{equation}
\label{equ:ColDdens13}
N(\textrm{total})_{\textrm{CO}} ^{13}=3.0\times10^{14} \frac{T\int{\tau ^{13}(\nu)d\nu}}{1-\exp{(-5.3/T)}}
\end{equation}

We used an assumed $^{13}$CO abundance to create a H$_2$ column density map. Commonly, the H$_2$ abundance relative to $^{13}$CO is quoted to be 3.6$\times$10$^{5}$ \citep{Frerking:1982,Bachiller:1986,Cernicharo:1984}. 
We assume a 50\% systematic error in this value, similar to the observed variation in $^{12}$CO abundance, and derive a range of masses based on a $^{13}$CO abundance of 1.4 to 4.2$\times$10$^{-6}$ relative to H$_2$.

\subsubsection{Mass Calculation}
H$_2$ column density maps were converted to maps of mass/pixel by $M_{\rm H_2} = 2 m_H A_{\rm pix} N_{\rm H_2} $, where $m_{\rm H}$ is the mass of a H atom, $A_{\rm pix}$ is the area of a fits file pixel and $N_{\rm H_2}$ is the average H$_2$ column density within a given region. The mass of a region could then be simply calculated by summing the corresponding pixels. This method is applied for both $^{12}$CO and $^{13}$CO-derived column densities. Due to sensitivity limitations, the $^{13}$CO-derived mass includes contributions from fewer map pixels than the $^{12}$CO-derived mass in all cases. Where clear $^{13}$CO emission is not present, mass calculations are not derived from $^{13}$CO images.

\section{Results}\label{sec:results}
Figures\,\ref{fig:Mom0s_1} and \ref{fig:Mom0s_2} display CO(1-0) components identified in this study, alongside $^{13}$CO(1-0) emission maps of corresponding velocity ranges. In the following sections, arranged by likely Galactic Arm Association (with reference to Figure\,\ref{fig:Spectra} and \ref{fig:PVplots}). Along the left side of Figures\,\ref{fig:Mom0s_1} and \ref{fig:Mom0s_2} are labels of Galactic arms corresponding to the headings of Sections\,\ref{sec:local} to \ref{sec:Norm}, which describe the features seen CO/$^{13}$CO(1-0) maps. For some clumpy structures present in maps which are argued to be a candidates for association with $\Jftf$ (see Section\,\ref{sec:GasAssoc}), we display spectral line parameters and derived physical parameters in Table\,\ref{tab:clumps}.

\begin{table*}
\begin{center}
\caption{Spectral line parameters and derived physical parameters for selected clump features. Velocity of peak ($v_{\rm LSR}$), line full-width half-maximum ($\fwhm$), peak H$_2$ column density calculated via two methods (X-factor method, \NX, and the $^{13}$CO optical depth correction method, \Nthir, respectively), mean $^{13}$CO(1-0) optical depth ($\tau ^{13}$), mean CO excitation temperature ($T$) and mass calculated via the X-factor method, M$^{12}$, and the $^{13}$CO optical depth correction method, M$^{13}$, are displayed. }\label{tab:clumps}
\begin{tabular}{|l|l|l|l|l|l|l|l|r|r|}
\hline 
Arm$^a$ & Region & \multicolumn{1}{l}{$v_{\rm LSR}$} & $\fwhm$  & \multicolumn{1}{l}{\NX} & \Nthir & $\tau ^{13}$ & $T$ & \multicolumn{1}{c}{M$^{X}$} &  M$^{13}$ \\
 		&  		& \multicolumn{2}{c|}{[$\kms$]}  & \multicolumn{2}{c|}{[$10 ^{21}\cmsqr$]}  & & [K] & \multicolumn{2}{l|}{\hspace{0.65cm}[\Msun]} \\
\hline
Sag.		&SNR$ ^b$		&	-	&	-	&0.5 - 1.5$^c$	&	-		&-	&	-	&	1700-5000	&	-\\
\hline
Scutum	&SNR$ ^b$		&	-	&	-	&0.3 - 1.0$^c$	&	-		& -	&-	&	19300-22000 &	-\\
-Crux	&$-40$A$ ^d$	&	$-$39.60$\pm$0.10		& 2.5$\pm$0.2	&2.4 - 7.2	&1.9 - 5.6	&0.06& 8	&	590-1700		&	100-300\\
(near)	&$-40$B$ ^d$		&	$-$38.97$\pm$0.03 	& 1.2$\pm$0.1	&2.2 - 6.4	&3.4 - 10	&0.2	& 10	&	220-670		&	120-370 \\
		&$-40$C$ ^d$		&	$-$36.73$\pm$0.06	& 2.8$\pm$0.2	&2.2 - 6.7	&3.7 - 11	&0.1	& 8 	&	2200-6800	&	690-2000\\
		&$-40$D$ ^d$		&	$-$42.49$\pm$0.03	& 2.2$\pm$0.1	&3.0 - 9.0	&3.3 - 9.6	&0.1& 9 	&	1100-3300	&	460-1400\\
		&$-50$A$ ^e$		&	$-$47.32$\pm$0.04	& 3.5$\pm$0.1	&4.1 - 12 	&6.4 - 19 	&0.1	&16 	&	3400-10000	& 	1800-5200\\
		&$-50$B$ ^e$	&	$-$47.18$\pm$0.02	& 2.0$\pm$0.1	&2.0 - 5.9		&2.0 - 5.9 	&0.2	&9 	&	1100-3200	&	520-1500\\
		&$-50$C$ ^e$	&	$-$49.70$\pm$0.10		& 3.4$\pm$0.1	&2.8 - 8.4 	&6.4 - 19	&0.1&12 	&	6300-19000	&	2000-5900\\
\hline
Norma	&SNR$ ^b$		&	-	&	-	&1.2 - 3.5$^c$	&	-		& -	&-	&	140000-420000&	- \\
-Cygnus	&				&		&		&				&			&	&	&				&	\\
\hline
\end{tabular}
\medskip\\
\end{center}
$^a$\textit{Most likely Galactic Arm association with reference to Figure\,\ref{fig:PVplots}. Distances used to derive masses were 1.2, 3.5 and 7.2\,kpc for the Sagittarius, Scutum-Crux and Norma-Cygnus arms, respectively. }
$^b$\textit{Elliptical region of $\Jftf$ identified at 843\,MHz, see Figure\,\ref{fig:HESS}. }
$^c$\textit{Average values within the region. }
$^d$	\textit{See Figure\,\ref{fig:Mom0s_1}. }
$^e$	\textit{See Figure\,\ref{fig:Mom0s_2}.}
\end{table*}

\subsection{Local gas}\label{sec:local}
In Figure\,\ref{fig:Mom0s_1}, CO(1-0) emission between $-$7.5 and $-$2.5$\kms$ is at a velocity consistent with local gas. Molecular gas clumps exist at the north and south of boundary of $\Jftf$, and outside the western boundary. With the exception of the western map boundary, minimal $^{13}$CO emission is seen. 

\subsection{The Sagittarius Arm}\label{sec:COsag}
CO(1-0)-traced molecular gas between $-$22 and $-$15\,$\kms$ is likely associated with the Sagittarius Arm at a distance of $\sim$1.2\,kpc. This gas is extended with diffuse components coincident with much of $\Jftf$, and has a total mass of 1.7-5.0\,$\Msun$ within the radio continuum boundary (shown in Figure\,\ref{fig:HESS}). Some $^{13}$CO emission is towards the south-east of the region.

\subsection{The Scutum-Crux Arm (far)}
CO(1-0)-traced molecular gas between $-$35 and $-$30\,$\kms$ is likely associated with the far-side of the Scutum-Crux Arm at a distance of $\sim$12\,kpc. The gas, which contains dense $^{13}$CO-traced clumps, is only observed towards the inner plane, to the north of $\Jftf$. 

\subsection{The Scutum-Crux Arm (near)}\label{sec:COscut}
CO(1-0)-traced molecular gas between $-$54 and $-$35\,$\kms$ is likely associated with the near-side of the Scutum-Crux Arm at a distance of $\sim$3.5\,kpc. Two velocity slices of this arm are displayed in Figure\,\ref{fig:Mom0s_1} (bottom) and \ref{fig:Mom0s_2} (top).

Between $-$44 and $-$35\,$\kms$, a series of approximately twelve 1 to 3\,arcminute CO(1-0)-traced clumps are present. These components are separated into regions labelled $-$40A, B, C and D in Figure\,\ref{fig:Mom0s_1}. Corresponding masses are on the order of 10$^{2}$ to 10$^3$\,$\Msun$ (see Table\,\ref{tab:clumps}), with most of the mass being either coincident or directly adjacent to the brightest gamma-ray component in the west. In Region\,$-$40C the molecular gas has an arc-like morphological distribution consisting of a north-south aligned chain of at least five clumps and a westward extension to the northern edge of the chain comprised of at least three clumps. Minimal diffuse extended CO(1-0) emission is observed across the central and south-east $\Jftf$ field.

The molecular mass in the $-$44 to $-$35\,$\kms$ velocity-range is separated from the Scutum-Crux arm in velocity-space, as indicated as ``$-$40B/C'' in Figure\,\ref{fig:PVplots}a. Along the adjacent Galactic arm, between $-$54 and $-$45\,$\kms$, a clump of mass $\sim$10$^{3}$\,$\Msun$ is coincident with the west of $\Jftf$, while clouds of mass $\sim$10$^{3}$\,$\Msun$ and $\sim$10$^{4}$\,$\Msun$ lie to the east and south, respectively. On average within each region, all Scutum-Crux Arm molecular gas components are observed to have $^{13}$CO(1-0) optical depth and CO excitation temperature within common ranges of 0.06 to 0.2 and 8 to 16\,K, respectively. 

Prompted by the existence of the distinct arc-like clumpy molecular structure, which is offset in velocity-space from the main arm, we examined atomic \HI\ emission from the Scutum-Crux arm for indications of an association with $\Jftf$. In Section\,\ref{sec:GasAssoc}, we discuss a candidate association for $\Jftf$. 

We note that a complete infrared bubble at [$l$,$b$]=[324.06,$-$0.81] \citep[S8,][]{Churchwell:2006} is coincident with region $-$40A to the north-east of $\Jftf$, suggesting a connection between high-mass OB star formation and the 10$^{2}$ to 10$^{3}$\,$\Msun$ clump. 
Alternatively, the Norma-Cygnus arm gas is also a potential association for the infrared bubble, particularly given the nearby broad and ring-like clump features in that region (see Figure\,\ref{fig:Mom0s_2}). It is unclear if this feature has any connection to the gamma-ray emission originating from the region.

\subsection{The Norma-Cygnus Arm}\label{sec:Norm}
CO(1-0)-traced molecular gas between $-$85 and $-$56\,$\kms$ is associated with the Norma-Cygnus Arm. The proximity to the tangent point makes precise distance determinations for this arm difficult (see Figure\,\ref{fig:PVplots}d). Assuming a distance of 7.2\,kpc, CO traces a total molecular mass of 1.4 to 4.2$\times 10 ^5$\,$\Msun$ towards $\Jftf$. In Figure\,\ref{fig:Mom0s_2}, this arm is divided into three velocity slices to highlight individual spectral features (images of Figure\,\ref{fig:Spectra} in rows labelled `Norma-Cygnus Arm').

Broad, spatially-extended CO/$^{13}$CO(1-0) emission features are coincident with the north-east of $\Jftf$, and a circular structure is centred on [$l$,$b$]=[324.05,$-$0.90] in the narrow $-$62.5 to $-$56\,$\kms$ velocity-range. Gas connected to these features extend towards the centre of the SNR, and gas from the north extends southwards coincident with the the northern half of the western TeV gamma-ray emission.Gas connected to this feature extends in velocity-space across 20\,km$^{-1}$, and can be seen in all three CO images of the Norma-Cygnus arm in Figure\,\ref{fig:Mom0s_2}. This, and the ring-like structure may indicate turbulence related to star formation.

\section{Discussion}\label{sec:Disc}
Towards the SNR $\Jftf$, Mopra CO(1-0) emission traces molecular gas from 5 Galactic arms. At least three of these have a molecular gas morphology which displays overlap with gamma-ray emission from the TeV gamma-ray shell. Such features offer the potential to explore possible hadronic components of gamma-ray emission in $\Jftf$. 

In Sections \ref{sec:COsag} to \ref{sec:Norm} CO(1-0)-traced molecular gas of the Sagittarius, Scutum-Crux and Norma-Cygnus arms has been described. Gas-gamma-ray overlap may make these arms candidates for association with $\Jftf$. 

In the case of the Scutum-Crux arm gas the molecular and atomic structure are suggestive of a wind-blown bubble candidate for $\Jftf$. In Section \ref{sec:GasAssoc} and \ref{sec:Modeling} we explore this scenario. 

We emphasise that the spectral line gas data presented in our study cannot be employed to explore scenarios where $\Jftf$ is not associated with gas. We investigate the potential of an association with gas that may take the form of a cavity blown out by a progenitor star \citep[e.g.][]{Fukui:2003,Fukuda:2014,Fukui:2017}, or overlap between gas clouds and gamma-ray emission \citep[e.g.][]{Aharonian:w28,Nicholas:2011,Abdalla:2016w49} which might suggest a hadronic scenario where CRs diffuse into nearby molecular gas. A perfect correlation between gas and gamma-rays is not necessary in the latter case, because hadronic gamma-ray emission will not only be proportional to the gas proton density, but the density of CRs. This is not expected to be uniform.

\subsection{A Scutum-Crux Arm Gas Association for HESS J1534-571}\label{sec:GasAssoc}

The discovery of clumpy molecular structures in Section\,\ref{sec:COscut} prompted a re-examination of SGPS $\HI$ data at nearby velocities (see Appendix\,\ref{app:HI}). Figure\,\ref{fig:Void} (top) highlights a dip in \HI\ emission at $\sim$47.5\,$\kms$, consistent with the Scutum-Crux arm at a distance of $\sim$3.5\,kpc. A $\HI$ gradient surrounds the SNR from the south-east to the north, then west. \citet{Abramowski:2017newshells} considered this dip as one of the two example potential ISM associations when discussing the energetics of $\Jftf$. To the south of $\Jftf$, where \HI\ emission is less pronounced, a $\sim$10$^3$ to 10$^4$\,$\Msun$ molecular cloud (``$-$50C'', see Figure\,\ref{fig:Mom0s_2} and Table\,\ref{tab:clumps}) is present, as traced by CO(1-0).
\begin{figure*}
\begin{center}
\includegraphics[width=0.65\textwidth]{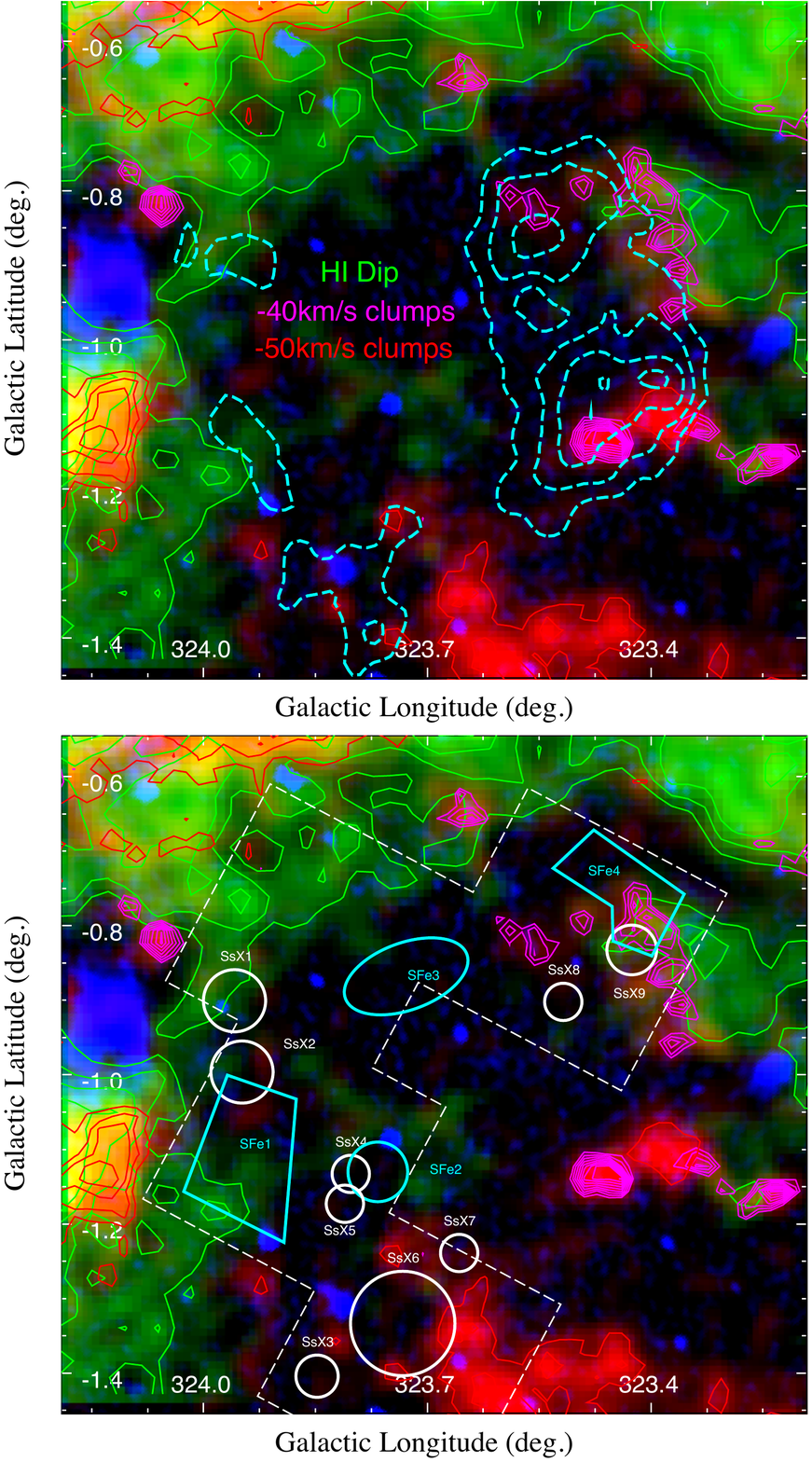}
\caption{SGPS \HI\ emission integrated from $-$50 to $-$45\,$\kms$ (green), Mopra CO(1-0) emission integrated from $-$54 to $-$45\,$\kms$ (red) and from $-$45 to $-$35\,$\kms$ (magenta contours) and MOST 843\,MHz continuum emission (blue). In the top image, HESS TeV gamma-ray significance (cyan contours) are overlaid. In the bottom image, X-ray emission regions from \citet{Saji:2018} are overlaid (see Section\,\ref{sec:GasAssoc} for details). The dashed-white region indicates Suzaku X-ray observation coverage, cyan regions indicate 6.4\,keV Fe emission line regions and white circles indicate soft X-ray emission regions. }\label{fig:Void}
\end{center}
\end{figure*}

The \HI\ dip may correspond to a bubble in atomic gas blown-out by the progenitor of $\Jftf$. In such a scenario analogous to that of SNR Vela\,Jr \citep{Fukui:2017,Maxted:2018_Vela} or $\RXJ$ \citep[e.g.][]{Inoue:2012}, nearby dense molecular clumps have sparse atomic outer layers stripped by strong stellar winds, leaving dense clumps within a wind-blown bubble. Following the supernova explosion, the ejecta expands into the void and interacts with the progenitor termination shock and stripped clumps. 

An association between $\Jftf$ and Scutum-Crux arm gas would suggest that the SNR was the product of a core-collapse (CC) event at a kinematic distance of $\sim$3.5\,kpc. The implied TeV luminosity of $\Jftf$ at this distance is comparable to SNRs $\RXJ$, $\VJ$, and $\Jsto$ (see Table\,\ref{tab:SNRs}).

\subsection{X-rays from $\Jftf$}
Signatures of an interaction might be detectable at X-ray energies. Dense molecular clumps embedded within young ($\sim 10^{3}$\,yr) SNR shells are sometimes associated with X-ray nodes caused by high energy electrons emitting non-thermally in magnetic fields on the boundary of dense gas. For example, X-ray emission peaks have been linked to shock compression regions near dense molecular gas in the young SNR $\RXJ$ \citep[e.g.][]{Cassam:2004,Acero:2009,Sano:2010,Inoue:2012,Maxted:2013rxj}, while thermal X-ray emission from plasma becomes prominent as SNRs approach the radiative phase of evolution, sometimes directly adjacent to molecular clouds \citep[e.g. see][]{Nakamura:2014}.

In Figure\,\ref{fig:Void} (bottom), we show Suzaku X-ray emission regions from \citet{Saji:2018} overlaid on the $\Jftf$ gas association candidate. Some of these regions encompass several peaks in X-ray emission. Four 6.4\,keV Fe line emission regions and nine soft X-ray emission regions were detected, labelled SFe1-4 (`Saji Fe') and SsX1-9 (`Saji soft X-rays'), respectively, in Figure\,\ref{fig:Void}. X-ray emission in the northwest (SFe4, SsX9) and south (SsX3, SsX5 and SsX9) appear to lie on the boundary of molecular clumps, suggesting that these originate from SNR shock-molecular cloud interactions with gas identified in this study. The strongest Fe emission detection, which is in the east of the remnant, SFe1, appears at the boundary of a large 3.4 to 10$\times$10$^3$\,$\Msun$ molecular cloud (region $-$50A), also suggestive of an association. However, no potential gas associations can be identified for X-ray emission regions SFe3, SFe2, SsX4, SsX5 and SsX8, while the coverage of the Suzaku maps do not include the southwest region of $\Jftf$, where $\sim$3 to 10$\times$10$^3$\,$\Msun$ of molecular gas lies. It follows that we have no firm conclusions on X-ray associations of $\Jftf$, and we leave further studies of this as future work.

\subsection{HESS J1534-571 Distance Constraints from Radio Brightness}\label{sec:Modeling}
In this section we investigate the $\Jftf$ distance through a comparison with models fitted to a sample of well-studied SNRs. 

\citet{Pavlovic:2014} compiled a $\Sigma-D$ calibrating sample of 65 Galactic SNRs with reliable distances, 1\,GHz flux measurements and angular diameters (required to calculate $\Sigma$ and $D$). The orthogonal fit to this data was then used as a calibrator to estimate the distance to SNRs with unreliable (or no) distance estimates. \citet{Vukotic:2014} then argued that a more robust calibration could be achieved with estimates of the data probability density function (PDF) in the $\log{\Sigma}-\log{D}$ plane instead of using the standard procedures based on fitted trend lines. This method was further statistically improved in \citet{Bozzetto:2017}, where the data were smoothed using a 2-D Gaussian to generate a probability density function (PDF). 
The kernel widths were obtained by minimizing the Bootstrap Integrated Mean Square Error \citep[i.e. BIMSE, for a detailed procedure description see][]{Bozzetto:2017}. The PDF was calculated on a $100\times100$ grid for the plotted variables (linear diameter and 1GHz radio continuum surface brightness, both expressed on logarithmic scales) using 100 bootstrap resamplings to yield a BIMSE value for each set of examined kernel bandwidths. The set of bandwidths with the smallest BIMSE value (0.16 in $\log{D}$ and 0.56 in $\log{\Sigma}$) was used for smoothing. 

In order to use the $\Sigma$-$D$ relation from \citet{Pavlovic:2014}, we scaled the MOST 843\,MHz flux measurements of $\Jftf$ to 1\,GHz assuming a simple scaling law $S\propto\nu^{\alpha}$ where the spectral index is $-0.5$. We note that a change of 20\% in the assumed spectral index results in a change in the 1\,GHz flux smaller than the resolution of the calculated PDF, thus errors from the extrapolation are considered negligible in our investigation. 
The MOST telescope recorded a $\Jftf$ 843\,MHz flux of $0.49$\,Jy, but this value excludes emission from structures at scales larger than 25\,arminutes \citep{Green:2014}. Following \citet{Abramowski:2017newshells}, we thus assume $0.49$\,Jy to be a lower limit. Double this value is assumed to be an appropriate upper limit, because most of the radio flux from SNRs generally originates from the forward and reverse shocks, which are likely narrower than 25\,arminutes in the case of $\Jftf$.

\begin{figure}
\includegraphics[width=0.5\textwidth]{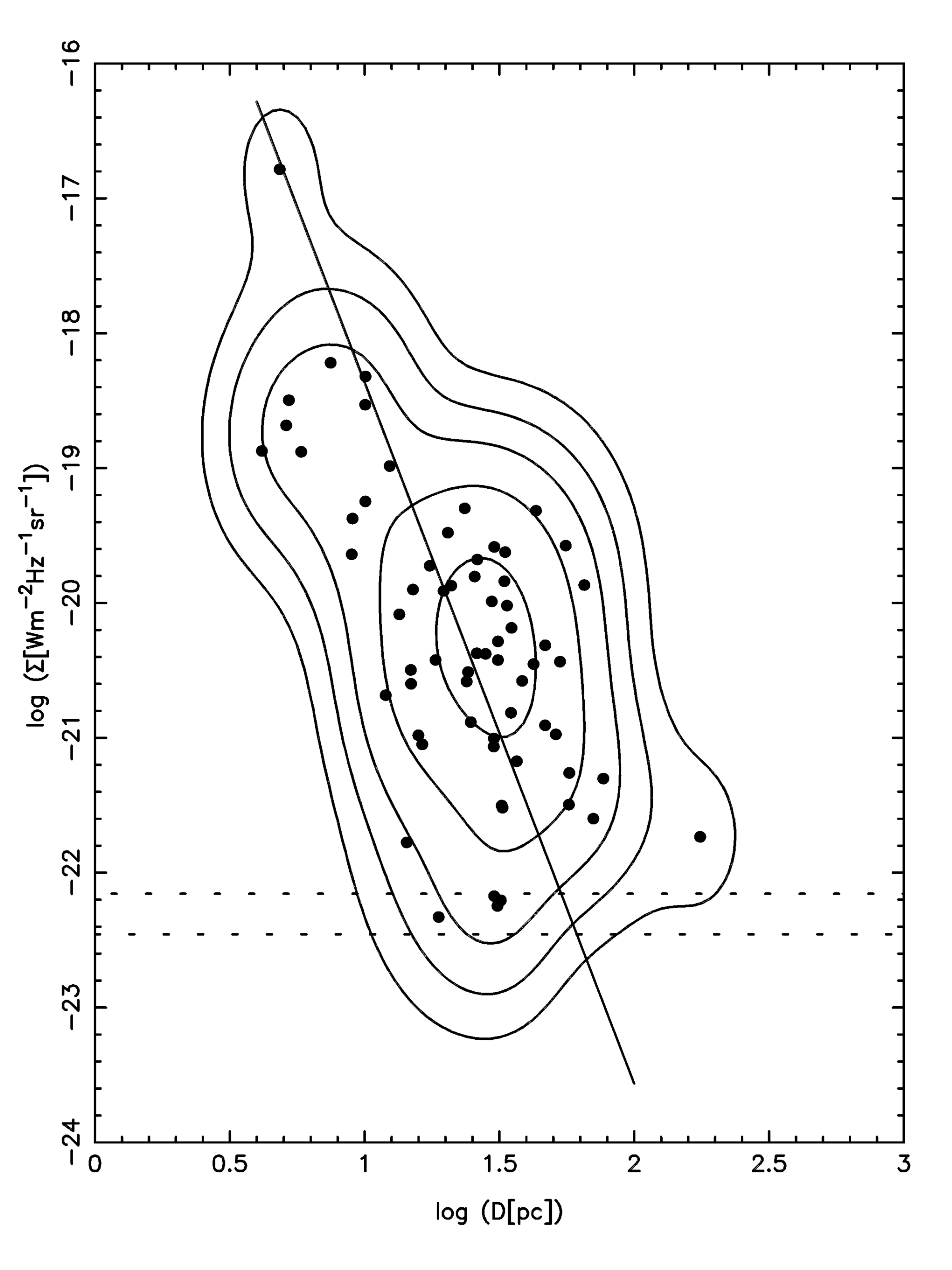}
\caption{The Probability Density Function of the 1\,GHz radio flux of 65 SNRs \citep[from][]{Pavlovic:2014}. The contour levels are at 0.02, 0.05, 0.1, 0.2 and 0.4. The solid line is the best fit from \citeauthor{Pavlovic:2014} while the dashed lines are the 1\,GHz flux bounds ($\Sigma$) extrapolated from the $\Jftf$ MOST 843\,MHz flux \citep{Green:2014}.}\label{fig:SigD}
\end{figure}

\begin{figure}
\includegraphics[width=0.48\textwidth]{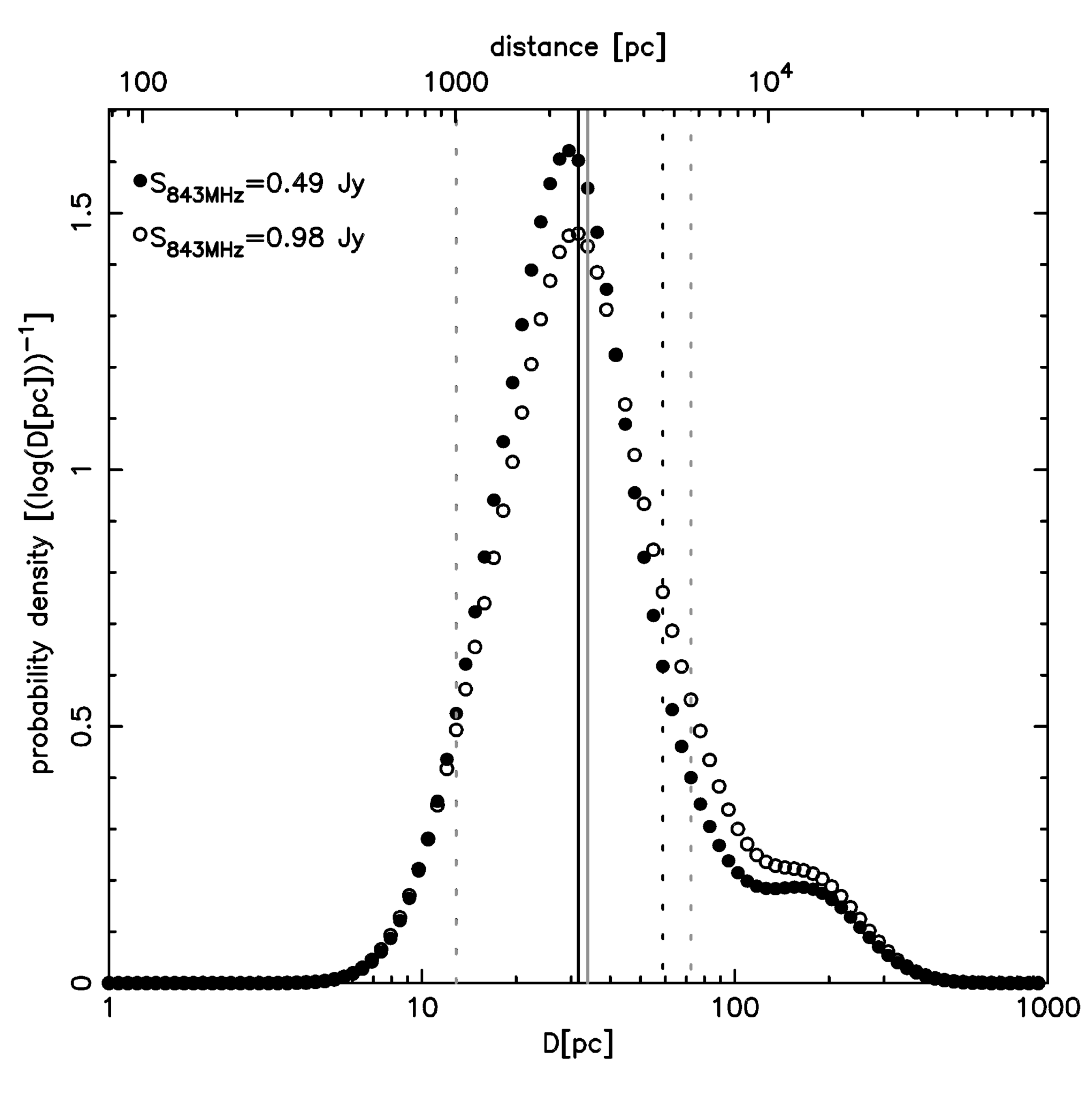}
\caption{The diameter probability density distributions derived from the PDF bidimensional distribution for the upper and lower $\Jftf$ 843\,MHz flux values from Figure\,\ref{fig:SigD} (open and closed circles, respectively) scaled to 1\,GHz assuming a spectral index of $-$0.5. The median values of the lower and upper flux-limit distributions are indicated by the solid and grey lines, respectively. Dashed vertical lines represent the $75\%$ confidence intervals of the two diameter distributions. The lower confidence interval (dashed grey) of the two diameter distributions is the same for both distributions, while the upper confidence intervals of the two cases are represented by the black (0.49\,Jy case) and grey (0.98\,Jy case) dashed lines.} \label{fig:PD}
\end{figure}

In Figure\,\ref{fig:SigD}, we display the PDF of the 1\,GHz radio continuum flux as a function of SNR diameter. The flux range of $\Jftf$ is indicated in this graph by 2 horizontal lines, and the probability distribution as a function of diameter along these 2 lines of constant surface brightness are displayed in Figure\,\ref{fig:PD}. The presented probability distributions are calculated by normalizing from bidimensional data in Figure\,\ref{fig:SigD} at the specified values. 

For the purpose of distance estimates (independent of any suggested gas association in this paper), we use median values of distributions in Figure\,\ref{fig:PD}. For the purpose of error estimates, we use a $75\%$ confidence interval by integrating to lower and higher values starting from the median value. The integration is performed in an assymetrical manner such that one side is integrated until the PDF value on that side falls below the PDF value on the other side \citep[for a more detailed description on error estimation algorithms in one dimensional PDFs see][]{Vukotic:2014}. 

The orthogonal fit calibration from \citet[][]{Pavlovic:2014} (see Figure\,\ref{fig:SigD}), $\log{\Sigma} = -13.16 - 5.2\log{D}$, yields a $\Jftf$ distance of 4.8\,kpc and 4.2\,kpc with fractional error $\sim$50\%, for lower and upper flux limits, respectively.\footnote{We note that the sample of SNRs used for the orthogonal fit line is biassed towards smaller diameters and larger radio continuum fluxes due to the sensitivity limits of radio continuum SNR searches, so the uncertainty may be larger.}

Our median-based estimate from Figure\,\ref{fig:PD} suggests a distance of 2.5\,kpc with a 1.0 to 4.6\,kpc confidence (75\%) interval and 2.6\,kpc with a 1.0 to 5.7\,kpc confidence interval, for the lower and upper flux values, respectively. These ranges cover smaller distances than those implied by the orthogonal offset method (Figure\,\ref{fig:SigD}, 4.2 to 4.8\,kpc $\pm$50\%), but both methods imply distances consistent with the near Scutum-Crux Arm gas ($3.5$\,kpc, see Section\,\ref{sec:GasAssoc}). This is in contrast to the Sagittarius and Norma-Cygnus arms at $\sim$1.2 and $\sim$7.2\,kpc, respectively.

We further note that our method generally yields lower distances than those implied by simple $\Sigma$-offset methods \citep[see][]{Pavlovic:2013,Pavlovic:2014}. 
The robust nature of the PDF-based method is likely to give a better insight into the confidence level of the estimates, which can also be very helpful when comparing with estimates from other (independent) methods - in this case possible CO associations.

In the case of $\Jftf$, \citet{Abramowski:2017newshells} also recognised that a $\Sigma$-offset distance ($\sim$15-20\,kpc) was flawed and suggested a distance closer to $\sim$5\,kpc based on the assumption of under-luminous radio emission, similar to $\RXJ$. Our calculation validates this suggestion and is consistent with the $3.5$\,kpc distance suggested by the candidate Scutum-Crux Arm gas association with $\Jftf$ and previous studies \citep[$\sim$4 to 5\,kpc and $\sim$6$\pm$2\,kpc,][respectively]{Araya:2017,Saji:2018}.

\subsection{The Age of HESS J1534-571}\label{sec:age}
In Section\,\ref{sec:GasAssoc}, we noted a plausible association of SNR $\Jftf$ with the Scutum-Crux Galactic arm at a distance of $\sim$3.5\,kpc. In this section, we discuss the age of the SNR using both a simple parametrisation that assumes similarities with $\RXJ$ and a more detailed model with varied initial inputs. 

A SNR age that mirrors the evolution of known TeV shells can be approximated by scaling known TeV shell-type SNR ages according to the Sedov relation,
\begin{equation}\label{equ:SedovRescale}
t\sim \left( \frac{R_{\rm sh}}{R_{\rm calib}}\right)^{5/2} t_{\rm calib}
\end{equation} where $R_{\rm sh}$ and $R_{\rm calib}$ are the shock radii of the SNR under scrutiny (in this case $\Jftf$) and the SNR used as a calibrator, respectively. $t_{calib}$ is the age of the SNR used as a calibrator. Implicit assumptions include both SNRs being in the Sedov phase, evolving in homogeneous media of the same density and an equivalent initial explosion energy. 
 
After 1600 years of evolution, SNR $\RXJ$ reached a diameter of $\sim$20\,pc \citep[e.g.][]{Wang:1997,Fukui:2003}\footnote{We note that some controversy around the 1600\,yr age solution for $\RXJ$ exists, but these can be rectified if the $\RXJ$ explosion was optically sub-luminous \citep[see][]{Fesen:2012}}, with a proportion of the SNR evolution time taking place within a wind-blown bubble. We use these values in Equation\,\ref{equ:SedovRescale} to approximate a $\Jftf$ age in the case that $\Jftf$ results from similar initial conditions to $\RXJ$. This yields ages of 1.1, 15 and 89\,kyr for assumed distances of 1.2, 3.5 and 7.2\,kpc, respectively. 

A second method estimated the $\Jftf$ age using SNR evolution modeling software by \citet{Leahy:2017}, which primarily utilises the approach outlined in \citet{Truelove:1999}. This software models the forward and reverse shock radii and velocity, and keV X-ray emission of SNRs as a function of SNR age. We considered three candidate distances (1.2, 3.5 and 7.2\,kpc). Model scenarios for both the 3.5\,kpc distance and the 1.2\,kpc distance were assumed to be low-explosion energy (0.5$\times$10$^{51}$\,erg) events due to low $\Jftf$ radio luminosity. The 7.2\,kpc distance model was assumed to have a moderate initial energy of 10$^{51}$\,erg. A homogeneous ISM density between 0.01 and 1.0\,cm$^{-3}$ was assumed in order to derive conservative age lower and upper limits, respectively, except in the case of the 3.5\,kpc distance model where we assume 0.01-0.1\,cm$^{-3}$. This is motivated by the existence of the suggested gas association for $\Jftf$ (see Figure\,\ref{fig:Void}, Section\,\ref{sec:GasAssoc}). All models were assumed to have an ejecta mass of 2\,M$_{\odot.}$. 

Our age solution for $\Jftf$ derived via \citeauthor{Leahy:2017} modeling are displayed in Table\,\ref{tab:SNRs}. For comparison, published age and distance constraints for other shell-type SNRs with shell-like gamma-ray morphology ($\VJ$, $\RXJ$ and $\Jsto$) are also included in Table\,\ref{tab:SNRs}. 

For the \HI\ dip distance of 3.5\,kpc, the $\Jftf$ would be in the Sedov phase of evolution with a corresponding age of 8 to 24\,kyr. The predicted forward shock velocity and electron temperatures are 400 to 1200\,km\,s$^{-1}$ and kT$\sim$0.2 to 1\,keV, respectively. Based on a plasma temperature derived from soft X-rays by \citet{Saji:2018}, kT$=$1.1$^{+0.5}_{-0.3}$\,keV, modelling possibly favours the ages nearer to 8\,kyr than 24\,kyr assuming that the X-ray detection primarily originates from electrons in the forward shock. The range of derived ages encompasses that estimated by scaling an $\RXJ$-like explosion (12\,kyr, Equation\,\ref{equ:SedovRescale}).

\begin{table*}
\caption{The diameter ($D$), age, coincident molecular mass, and 1 to 10\,TeV spectral index ($\Gamma _{\gamma}$) and luminosity ($L_{\gamma,1-10\,TeV}$) \citep[with reference to][]{Abramowski:2017newshells} of core-collapse TeV supernova remnants with constrained distances ($d$). Where multiple possible distance solutions exist, each distance is considered.}\label{tab:SNRs}
\begin{center}
\begin{tabular}{|l|r|r|r|r|r|r|}
\hline
SNR		&	$d^{a}$ 	& $D^{a}$  		& Age$^{a}$				& Molecular				& $\Gamma _{\gamma}$	& $L_{\gamma,1-10\,TeV}$ \\
		&	(kpc)	& (pc) 			& (yr)					& Mass$^{a}$ ($\Msun$)	 	&			&(10$^{33}$\,erg\,s$^{-1}$)	\\
\hline
$\VJ$	&	0.75 	& 26				& $\sim$	10$^{3}$			& $\sim$1$\times 10 ^{3}$	&	1.81		&	5.7		\\
\hline
$\RXJ$	&	1.0		& 20				& 1.6$\times$10$^{3}$	& 9$\times 10 ^{3}$			&	2.06		&	7.2		\\
\hline
$\Jsto$	&	3.2		& 30				& $\sim$2.5$\times$10$^{3}$	& 5$\times 10 ^{4}$			&	2.32 	&	8.5		\\
&	5.2	& 49		& $~~~~~^{\prime\prime}$	& 4-5$\times 10 ^{4}$		&			&	22.4		\\
\hline
$\Jftf$	&	1.2		& 17				& $^{b,e}$1.3-5.5$\times$10$^{3}$	& 1.7-5.0$\times 10 ^{3}$	&		&	1.1		\\	
&	3.5	& 49		& $^{c,e}$8-24$\times$10$^{3}$	& 0.9-2.2$\times 10 ^{4}$	&	2.51			&	9.6		\\
&	7.2	& 100	& $^{d,e}>$32$\times$10$^{3}$	& 1.4-4.2$\times 10 ^{5}$	&			&	40.4		\\
\hline
\end{tabular}\\
\medskip
\end{center}
\textit{$^{a}$Gas associations, age and distance constraints: \citep[\VJ:][]{Katsuda:2008,Allen:2015,Fukui:2017}, 
\citep[$\RXJ$:][]{Wang:1997,Fukui:2003,Moriguchi:2005,Aharonian:2007rxj,Fukui:2012}, \citep[$\Jsto$:][]{Tian:2008,Fukuda:2014,Cui:2016,Maxted:2018_HESSJ1731}}\\
\textit{$^{b}$Assuming an initial explosion energy of 0.5$\times$10$^{51}$\,erg and constant H density of 0.01-1.0\,cm$^{-3}$.}
\textit{$^{c}$Assuming an initial explosion energy of 0.5$\times$10$^{51}$\,erg and constant H density of 0.01-0.1\,cm$^{-3}$.}
\textit{$^{d}$Assuming an initial explosion energy of 1.0$\times$10$^{51}$\,erg and constant H density of $>$0.01\,cm$^{-3}$.}
\textit{$^{e}$See Section\,\ref{sec:Modeling} for further details.}
\end{table*}

\subsection{Future Studies}\label{sec:future}
In previous SNR studies, the identification of X-ray peaks adjacent to dense clumps have been noted to signal a gas-shock interaction \citep[e.g.][]{Cassam:2004}, while the comparison of spectral-line-derived column densities with X-ray absorption column density measurements \citep[e.g.][]{Maxted:2018_HESSJ1731} have constrained SNR distance. In this context, further X-ray emission observations that achieve full spatial-coverage of $\Jftf$ will be a key step to understanding the distance, age and nature of $\Jftf$, particularly towards regions $-$40D and $-$50B, which currently lack coverage at keV energies. Furthermore, additional X-ray coverage will aid in a search for an associated CCO, which could constrain the SNR distance and age with modeling of thermal emission.

Furthermore, in the bright TeV core-collapse SNRs $\RXJ$, $\Jsto$ and $\VJ$, gamma-ray features may hint at the existence of CR acceleration through spatial gas/gamma-ray correspondence within the TeV shell \citep[][respectively]{Moriguchi:2005,Fukuda:2014,Fukui:2017}, CR interactions in nearby gas \cite[e.g. potentially][]{Cui:2016,Capasso:2017,Maxted:2018_HESSJ1731}, or spectral flattening resulting from energy-dependent hadron diffusion into dense molecular cores \citep{Gabici:2009,Zirakashvili:2010,Inoue:2012,Fukui:2012,Maxted:2012,Gabici:2014}. The $\sim$10-fold sensitivity-increase and arcminute-scale resolution of the next generation of gamma-ray instrument, Cherenkov Telescope Array \citep[CTA,][]{Acero:2017}, will allow CR acceleration in young SNRs to be probed by searching for these gas/gamma-ray signatures. In the case of $\Jftf$, sub-arcminute Mopra CO/$^{13}$CO(1-0) maps will enable investigations of features in high resolution gamma-ray maps to help distinguish between hadronic and leptonic emission scenarios.

\section{Conclusions}
We present Mopra Southern Galactic Plane CO and $^{13}$CO(1-0) emission maps towards the TeV gamma-ray supernova remnant $\Jftf$ (G323.). We examine the morphology of molecular cloud structures in five Galactic arms, and highlight Scutum-Crux arm molecular cores that are at a consistent velocity with a \HI\ dip. We suggest this to be associated with a progenitor wind-blown cavity of a $\Jftf$ core-collapse event. The distance corresponding to the cavity, $\sim$3.5\,kpc, is consistent with our two independent analyses of the radio continuum brightness ($\sim$4.5$\pm$2.5\,kpc and $\sim$2.5$^{+3.0}_{-1.5}$\,kpc) and recently-published distance estimates from leptonic gamma-ray emission modelling and X-ray absorption calculations ($\sim$4 to 5\,kpc and $\sim$6$\pm$2\,kpc, respectively). Based on the assumption of a 3.5\,kpc distance solution, we suggest the SNR to be in the Sedov-Taylor phase of evolution with an age of 8 to 24\,kyr.

\section*{Acknowledgments} 
The Mopra Telescope is part of the Australia Telescope and is funded by the Commonwealth of Australia for operation as a National Facility managed by the CSIRO. The University of New South Wales Mopra Spectrometer Digital Filter Bank used for these Mopra observations was provided with support from the Australian Research Council, together with the University of New South Wales, the University of Adelaide, University of Sydney, Monash University and the CSIRO. We thank the Australian Research Council for helping to fund this work through a Linkage Infrastructure, Equipment and Facilities (LIEF) grant (LE16010094). D. Urosevic, B. Vukotic and M. Z. Pavlovic acknowledge support by the project No. 176005, ``Emission nebulae: structure and evolution,'' supported by the Ministry of Education, Science, and Technological Development of the Republic of Serbia. We thank Shigetaka Saji for providing us with Suzaku X-ray emission coordinates promptly following our request.




\bibliographystyle{mnras}
\bibliography{ReferencesHESSJ1534} 




\appendix
\section{HI investigation}\label{app:HI}
We examined SGPS \HI\ \citep{McClure:2005} emission towards the supernova remnant, $\Jftf$. \HI\ emission was integrated in 5$\kms$ velocity slices in a search for \HI\ dip features that align with the MOST 843\,MHz radio emission of $\Jftf$, while colour scales were adjusted to highlight \HI\ emission gradients. A \HI\ dip identified in this study is highlighted in Figure\,\ref{fig:Void}.

\begin{figure*}
\includegraphics[width=1.0\textwidth]{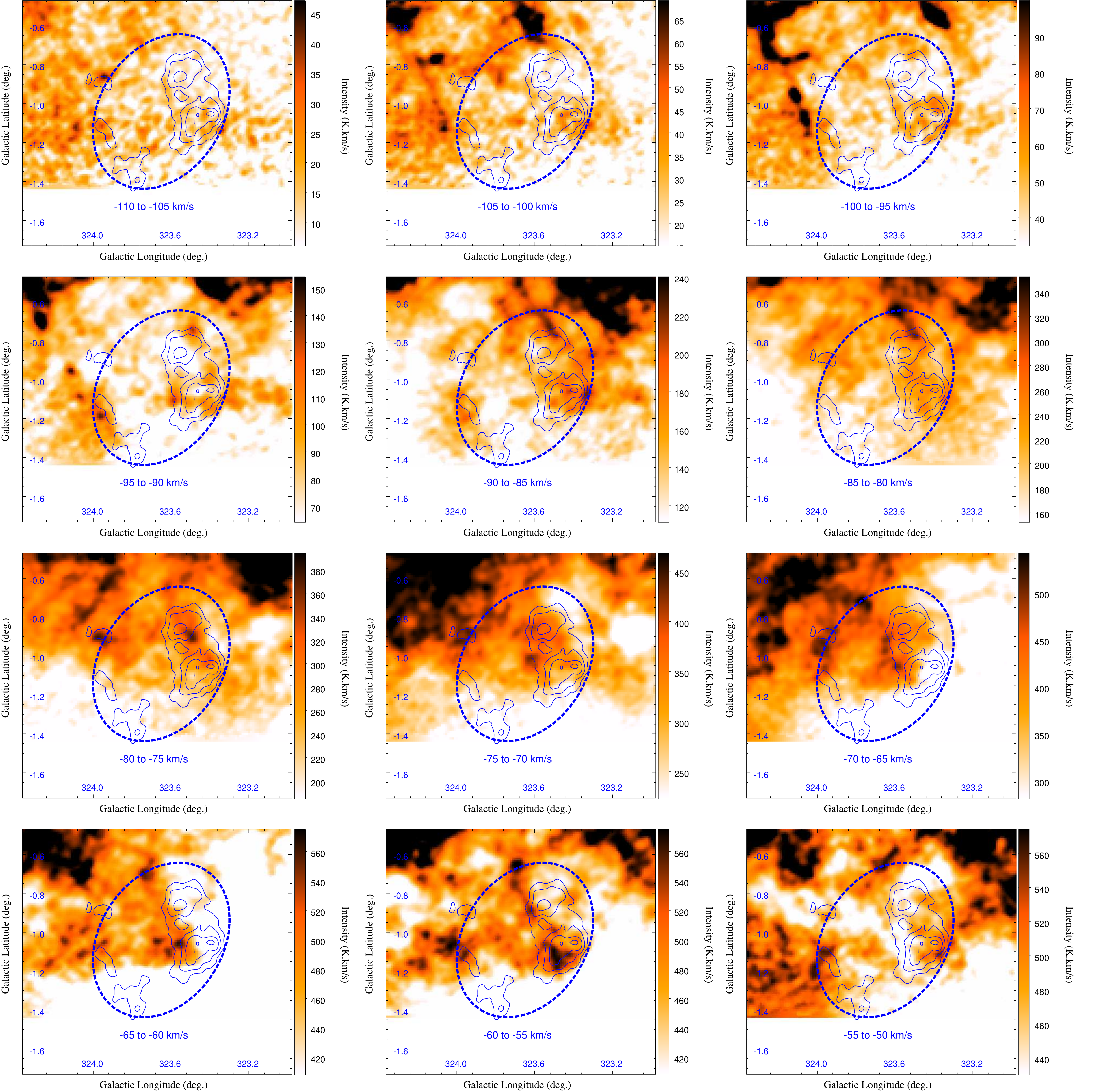}
\caption{HI emission integrated in 5\,km/s velocity slices. 3, 4, 5 and 6$\sigma$ TeV gamma-ray emission contours of the associated source $\Jftf$ are overlaid. A blue dashed ellipse indicates the characterisation of the 843\,MHz emission, which has a centre of [$\alpha$,$\delta$]=[05:34:30.1$-$57:12:03] and axes diameter of 51$\times$38$^{\prime}$. \label{fig:HIimages1} }
\end{figure*}
\begin{figure*}
\includegraphics[width=1.0\textwidth]{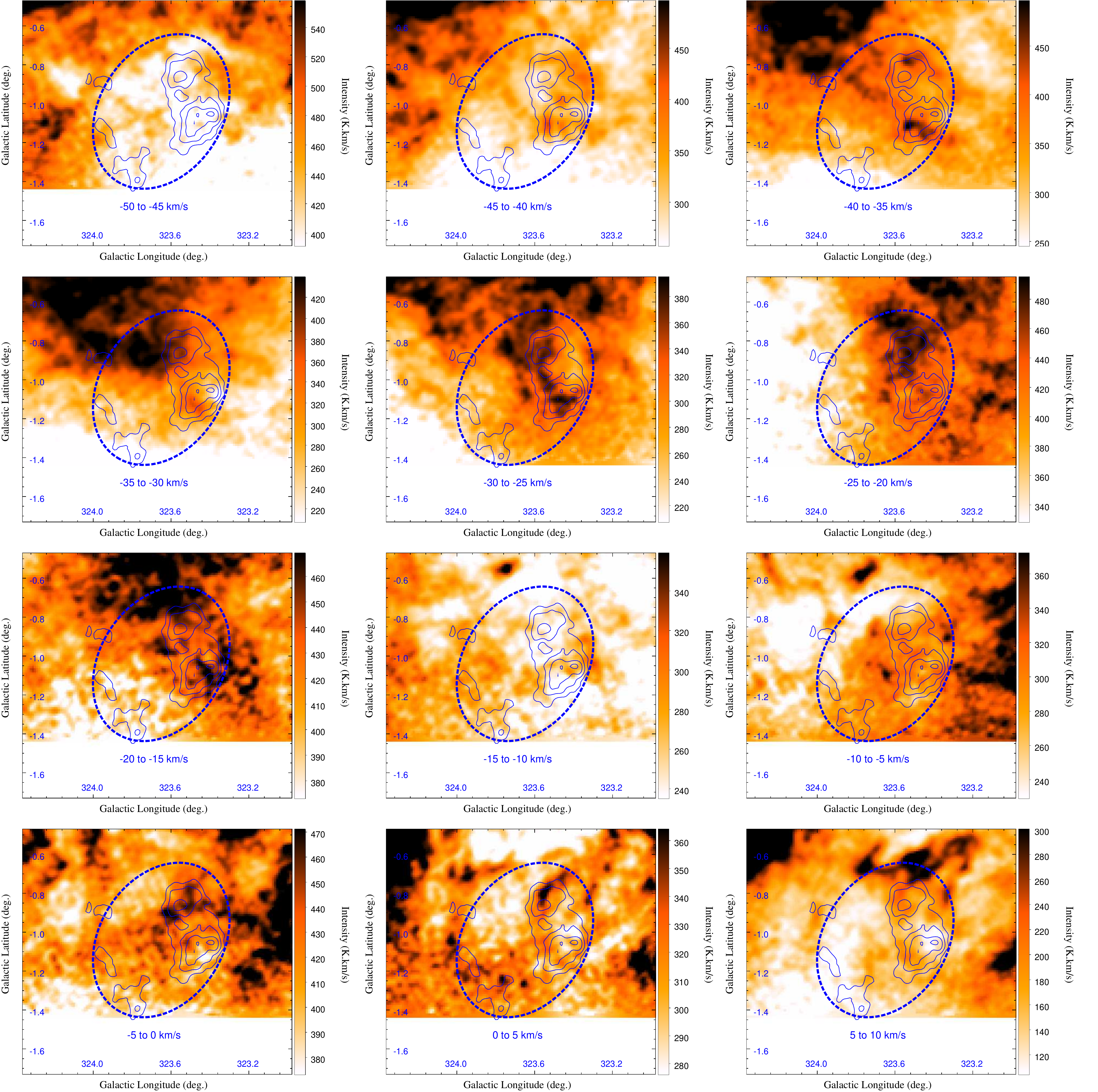}
\caption{Same as Figure\,\ref{fig:HIimages1}, but for different velocity-integration ranges. \label{fig:HIimages2} }
\end{figure*}


\bsp	
\label{lastpage}
\end{document}